\documentclass[11pt,a4paper]{article}

\usepackage[T1]{fontenc}
\usepackage[utf8]{inputenc}
\usepackage{lmodern}
\usepackage{microtype}

\usepackage[margin=1in]{geometry}

\usepackage{graphicx}
\usepackage{float}

\usepackage{caption}
\usepackage{subcaption}

\usepackage{amsmath}

\usepackage[numbers,sort&compress]{natbib}

\usepackage[hidelinks,
            colorlinks=true,
            linkcolor=blue,
            citecolor=blue,
            urlcolor=blue]{hyperref}

\usepackage{authblk}

\usepackage{xspace}
\usepackage{booktabs}
\usepackage{enumitem}

\title{e112: A Context-Aware Mobile Emergency Communication Platform Leveraging Smartphone Sensing and Cloud Services%
\thanks{This paper was presented at IDRC~2025: The 8th International
Disaster and Risk Conference.}}

\author[1]{Katerina Ioannidou}
\author[2]{Marios D.\ Dikaiakos\thanks{Corresponding author. Email: \href{mailto:mdd@ucy.ac.cy}{mdd@ucy.ac.cy}}}
\author[3]{Athena Stassopoulou}

\affil[1]{Department of Information Technology, Uppsala University,
          SE-751~05 Uppsala, Sweden}
\affil[2]{Department of Computer Science, University of Cyprus,
          Nicosia 2109, Cyprus}
\affil[3]{Department of Computer Science, University of Nicosia,
          Nicosia 2417, Cyprus}

\date{}   

\begin{document}

\maketitle

\begin{abstract}
This paper presents e112, a context-aware mobile emergency response application
designed to strengthen communication between citizens and authorities during
disasters. Building on the ubiquity of smartphones, the system provides SOS
requests, incident reporting, customized alerts, evacuation guidance, and
moderated community interaction, supported by a cloud-based back end and an
operator dashboard for situational awareness. A user-centered design approach
guided our development, ensuring clarity and usability under stressful
conditions. Evaluation through usability studies and technical audits
demonstrated high user satisfaction, robust performance, and accessibility. The
results show that a simple, well-designed mobile application can significantly
enhance emergency preparedness and response, reducing risks to human life during
climate change--driven emergencies.
\end{abstract}


\section{Introduction}
\label{sec:intro}

The frequency, severity, and impact of natural disasters have risen sharply in
recent years, leading to loss of life, environmental degradation, urban
infrastructure damage, and significant economic
costs~\citep{Tin:EMDATAnalysis:2024,CRED:EMDAT2024:2025}. Climate change is a
central driver, producing higher temperatures, prolonged droughts, and more
frequent extreme weather events~\citep{IPCC:2022}. These conditions accelerate
wildfires that devastate ecosystems and human
settlements~\citep{wildfires:essd:2024}, while leaving burned landscapes highly
vulnerable to subsequent floods and landslides. Rapid urbanization further
amplifies risks from both natural and human-induced hazards. Crucially, many
disasters arise from compound events, where interacting climate drivers converge
across spatial and temporal scales, generating cascading and magnified
effects~\citep{Zscheischler:ClimateChange:2018}.

Minimizing risk to human life during fast-moving disasters depends on the
effectiveness of emergency services that can be alerted instantly, provide clear
protective guidance, integrate situational awareness, and coordinate rapid
deployment of assistance to the most affected areas. To support these functions,
authorities (civil protection, fire brigade, police) operate emergency call
centers as core components of \emph{Emergency Services Communication Systems}
(ESCS). Accessible through standardized numbers---112 in the EU, 911 in the US
and Canada, and 999 in the UK---these centers ensure universal access to first
responders. Emergency lines primarily enable citizens to report incidents and
request aid, while also allowing authorities to disseminate urgent instructions
such as evacuation orders via SMS and other alerting mechanisms.

Given the ubiquity of smartphones and mobile applications, these technology
platforms present a significant opportunity to advance emergency communication
and strengthen public safety services. On the one hand, smartphones can leverage
their embedded sensors (GPS, cameras, and microphones) to capture critical
contextual data, including precise geographic location and multimedia evidence,
thereby enhancing, validating, and partially automating the information that
individuals report to authorities during emergencies. On the other hand,
authorities can disseminate targeted, timely, and context-aware alerts that
range from evacuation routes and hazard warnings to locations of safe areas and
available resources, thereby improving situational awareness, supporting safe
and effective evacuation from danger zones, and enhancing coordinated emergency
response.

As authorities worldwide modernize traditional ESCS frameworks, transitioning
from circuit-switched infrastructures to Internet Protocol (IP)-based
platforms~\citep{wikipedia:NG911,jordan:escs:mascots24}, smartphones and mobile
applications are expected to emerge as the primary interface for two-way
communication between citizens and emergency services. In this paper, we present
the design, implementation, and evaluation of e112, a prototype interactive
smartphone application platform developed to provide rapid and effective
assistance during emergency situations. The application addresses key limitations
of traditional communication methods by enabling structured text- and
multimedia-based interaction with emergency operators. Core features include an
SOS button for immediate help requests, real-time location tracking, and
multimedia uploads. In addition, e112 delivers timely local updates and supports
on-demand, location-aware private and group chat to facilitate information
exchange and coordination between volunteers and emergency services.

The remainder of this paper is organized as follows: Section~\ref{sec:related}
discusses related work; Section~\ref{sec:concept} presents the main concepts
implemented in the application; Section~\ref{sec:arch} presents the software
design and server components and discusses implementation aspects; evaluation of
the prototype is presented in Section~\ref{sec:eval}; and we conclude in
Section~\ref{sec:concl} with a summary of conclusions and future work.

\section{Background and Related Work}
\label{sec:related}

Since the earliest developments of public warning systems, authorities have
sought to design alerts and delivery mechanisms that balance accuracy, clarity,
and urgency~\citep{NASEM:EAW:2018}. As these systems evolve from traditional
broadcast methods to mobile communications and social platforms, scientific
studies increasingly explore how message features and communication dynamics
could influence public response during
crises~\citep{Chen:EmergencyMessagingInsights:2024}. In an early study of this
question, the authors of~\citep{Mileti:WRM:1990} reviewed a large number of
prior studies, identified the stages of a typical warning sequence (hear
$\to$ confirm $\to$ understand $\to$ believe $\to$ personalize $\to$ act) and
demonstrated its applicability to both pre- and post-impact
contexts~\citep{MiletiOBrien:SocialProblems:1992}. Their Warning Response Model
(WRM) stressed that effective warnings must be hazard-specific, provide clear
protective guidance, indicate who is at risk and where, state when action is
required, and identify trusted sources. They also showed that individuals rarely
act on the first warning alone but instead seek confirmation from multiple
channels and social contacts, making clarity, consistency, and repetition across
sources essential for eliciting timely and appropriate protective action. They
posited that the response of the public to post-impact warnings is shaped not
only by message content but also by the recipients' prior experiences of loss,
damage, and disruption. Authorities have sought to operationalize these
principles in the design of systems like the Emergency Alert System
(EAS)~\citep{FCC:EAS}, and adapt them to the mobile era. Typically, current EAS
employ \emph{Wireless Emergency Alerts}
(WEAs)~\citep{WEAEnhancements:FEMA:2025}, which are delivered directly to
personal communication devices in the form of short 90-character messages issued
by government authorities and delivered free of charge by wireless carriers to
individuals who face imminent threats such as fires, tornadoes, or floods. A
recent study explored what the addition of maps to WEAs could bring from the
perspective of decision clarity, compliance to instructions, and information
sharing in various disaster scenarios~\citep{liu:prr:2017}. This study presented
two experiments involving U.S.\ adults and found that while maps marginally
improved message comprehension, their overall impact on compliance and sharing
behavior was limited. This result highlights the need for further research on
optimizing visual aids in emergency messaging, for instance by dynamic and
interactive map visualizations~\citep{Bartling:MobileMapContext:2022} and
support for real-time decision-making.

Another key factor in the effectiveness of emergency communication systems are
offline social networks, such as family, community groups, kinship ties, and
voluntary associations. These networks strongly influence how warnings are
shared, interpreted, and acted
upon~\citep{Mileti:WRM:1990,MiletiOBrien:SocialProblems:1992}. Close social
connections not only facilitate the dissemination and discussion of warnings but
also strengthen understanding, credibility, and timely response. The mutual
trust embedded in such networks enhances the persuasiveness of warnings and
motivates protective action, while also providing essential support and
resources during crises. Consequently, as online social networks have become a
ubiquitous platform for communication and rich information
exchange~\citep{Pallis:OSN:2011}, new opportunities have emerged for
disseminating emergency alerts~\citep{Facebook:CrisisResponse,VeraBurgos2020,%
Atkinson:FacebookCrisis:2021} and embedding social communication features into
alert systems. However, effective use of these platforms requires sensitivity to
cultural and community contexts that shape how warnings are perceived and acted
upon. Moreover, while social networks can accelerate the spread of accurate
information, they are also vulnerable to inadvertent misinformation or
deliberate disinformation attacks~\citep{Paschalides:OSNM:2021}. Ensuring the
accuracy and consistency of emergency messages is therefore essential to prevent
confusion and preserve the effectiveness of warning systems.

\section{Main Design Concepts}
\label{sec:concept}

Based on the above observations we implemented the e112 system as a
cross-platform smartphone application integrated with a cloud-based back end.
The mobile application provides citizens with access to location-aware emergency
reporting and disaster-related information. The back end supports emergency
operators through a web-based dashboard, manages context-aware communication
with users, and integrates external services such as Google Maps for geospatial
data, Firebase~\citep{Firebase:2025} for real-time synchronization, and
Twilio~\citep{Twilio:Platform:2025} for secure user verification.

\begin{figure*}[t]
\centering
\includegraphics[width=\textwidth]{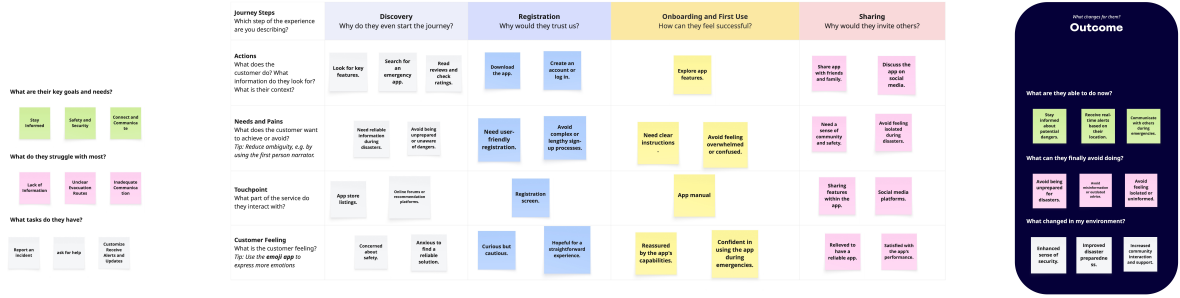}
\caption{Customer-journey map for user registration and application on-boarding.}
\label{fig:customerJourney}
\end{figure*}

\begin{figure*}[t]
\centering
\includegraphics[width=\textwidth]{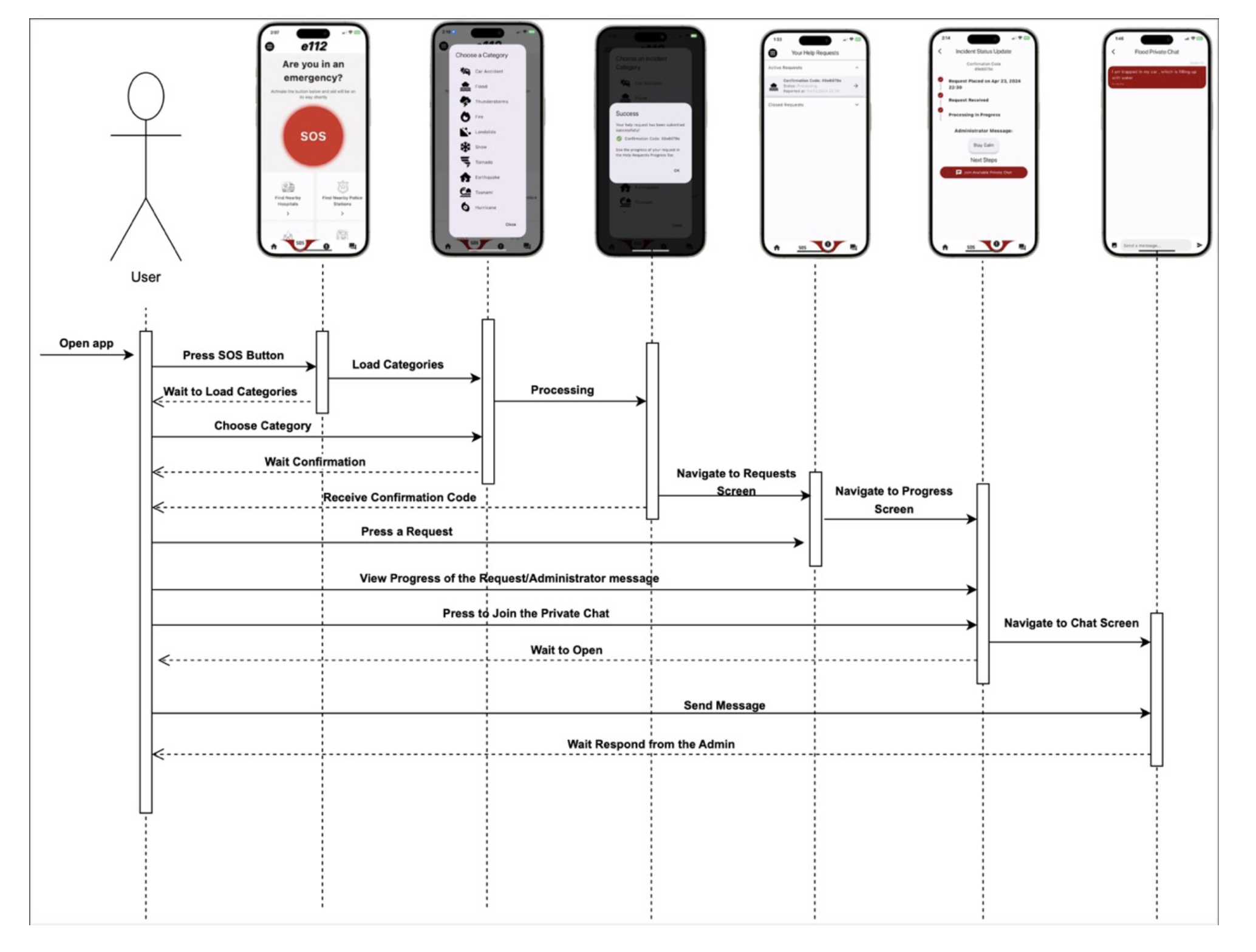}
\caption{UML Interaction diagram derived from customer-journey analysis of a
use-case where a user in danger seeks help through the e112 app; the interaction
diagram (bottom) is mapped to the respective app's GUI screens (top).}
\label{fig:sosCase}
\end{figure*}

We formulated the functionality of the mobile application by designing first the
\emph{User Experience} (UX) it should offer. UX addresses the overall experience
of a person interacting with a service, including usability, utility, and
emotional impact, with the goal of solving real user problems effectively and
intuitively~\citep{Klein:UXLean:2013}. For the UX analysis, we adopted a
user-centered design methodology inspired by design thinking principles, and
modeled user needs through \emph{personas} (end users and system operators),
\emph{customer journeys}, and \emph{empathy maps}. Customer journey
design~\citep{FolstadKvale:CustomerJourneys:2018} provides a structured view of
the stages users go through, highlighting critical touch-points for the design
of the application's interactivity (see Figure~\ref{fig:customerJourney}). The
purpose of empathy maps is to capture what users see, hear, think, feel, say,
and do at each stage of application use, thereby informing the design process by
revealing their underlying motivations, frustrations, and
needs~\citep{LewrickLinkLeifer:2018}. For example, in a wildfire scenario,
journey mapping clarified the sequence of actions (receiving an alert, navigating
to a shelter), while empathy mapping revealed stress, confusion, and the need
for reassurance that could hinder decision-making. We consolidated the
customer-journey analysis in UML interaction diagrams (see
Figure~\ref{fig:sosCase}).

\begin{figure}[t]
  \centering
  \begin{minipage}{0.32\linewidth}\centering
    \includegraphics[width=\linewidth]{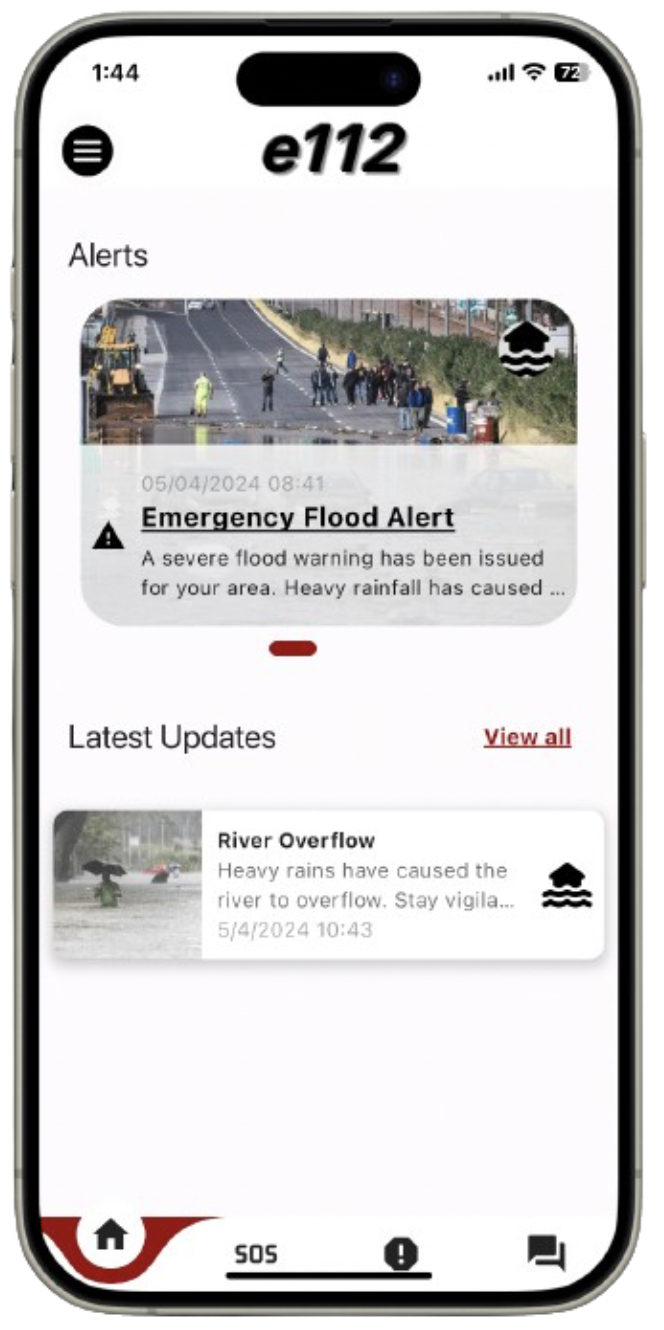}\\[-2pt]
    \footnotesize (a)
  \end{minipage}\hfill
  \begin{minipage}{0.32\linewidth}\centering
    \includegraphics[width=\linewidth]{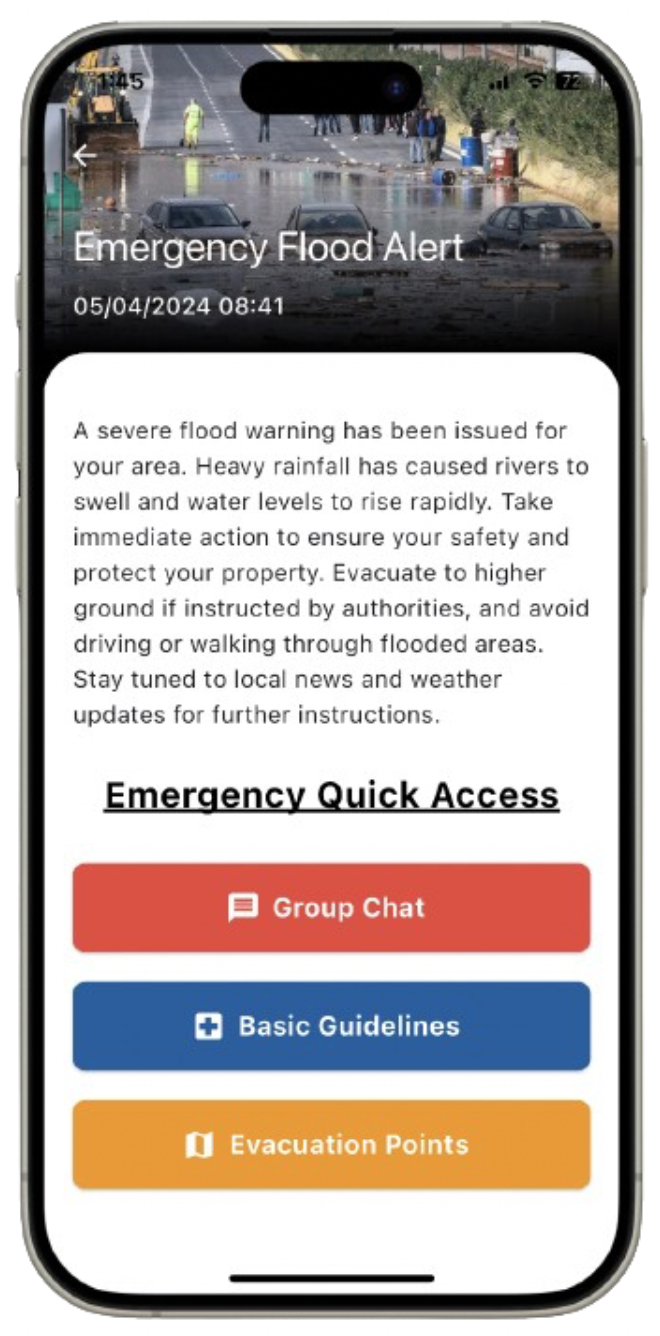}\\[-2pt]
    \footnotesize (b)
  \end{minipage}\hfill
  \begin{minipage}{0.32\linewidth}\centering
    \includegraphics[width=\linewidth]{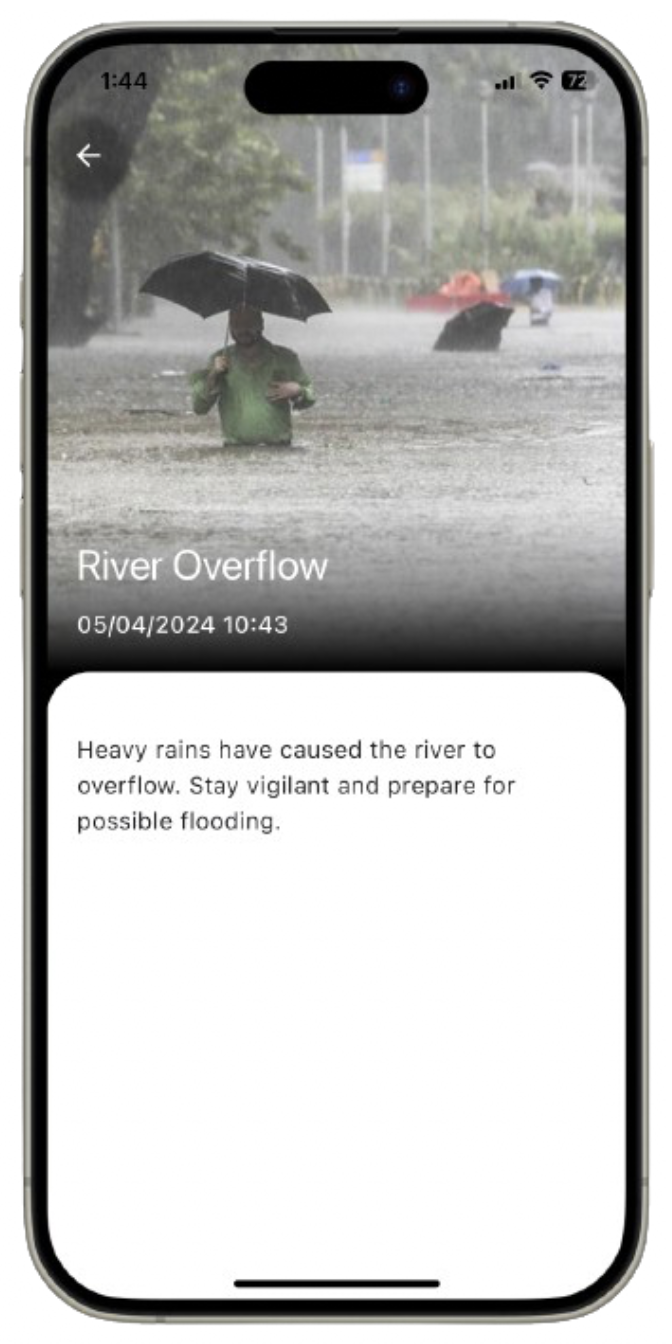}\\[-2pt]
    \footnotesize (c)
  \end{minipage}
  \vspace{4pt}
  \begin{minipage}{0.32\linewidth}\centering
    \includegraphics[width=\linewidth]{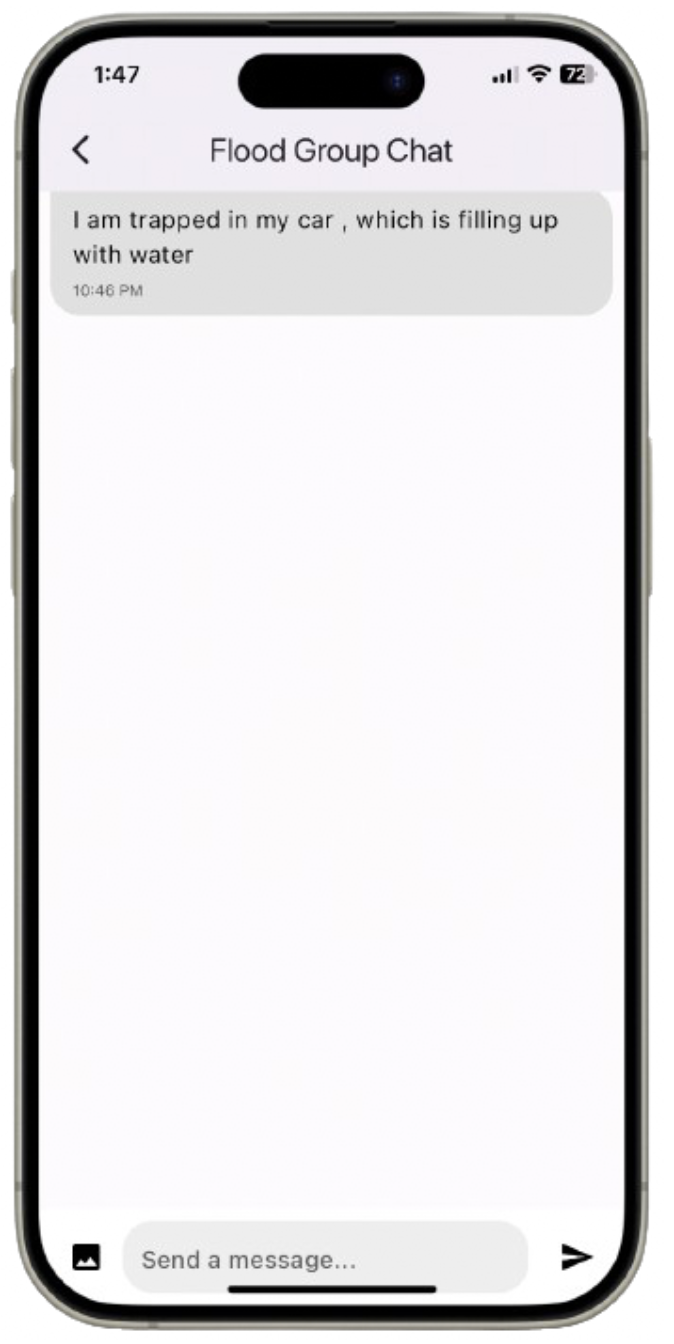}\\[-2pt]
    \footnotesize (d)
  \end{minipage}\hfill
  \begin{minipage}{0.32\linewidth}\centering
    \includegraphics[width=\linewidth]{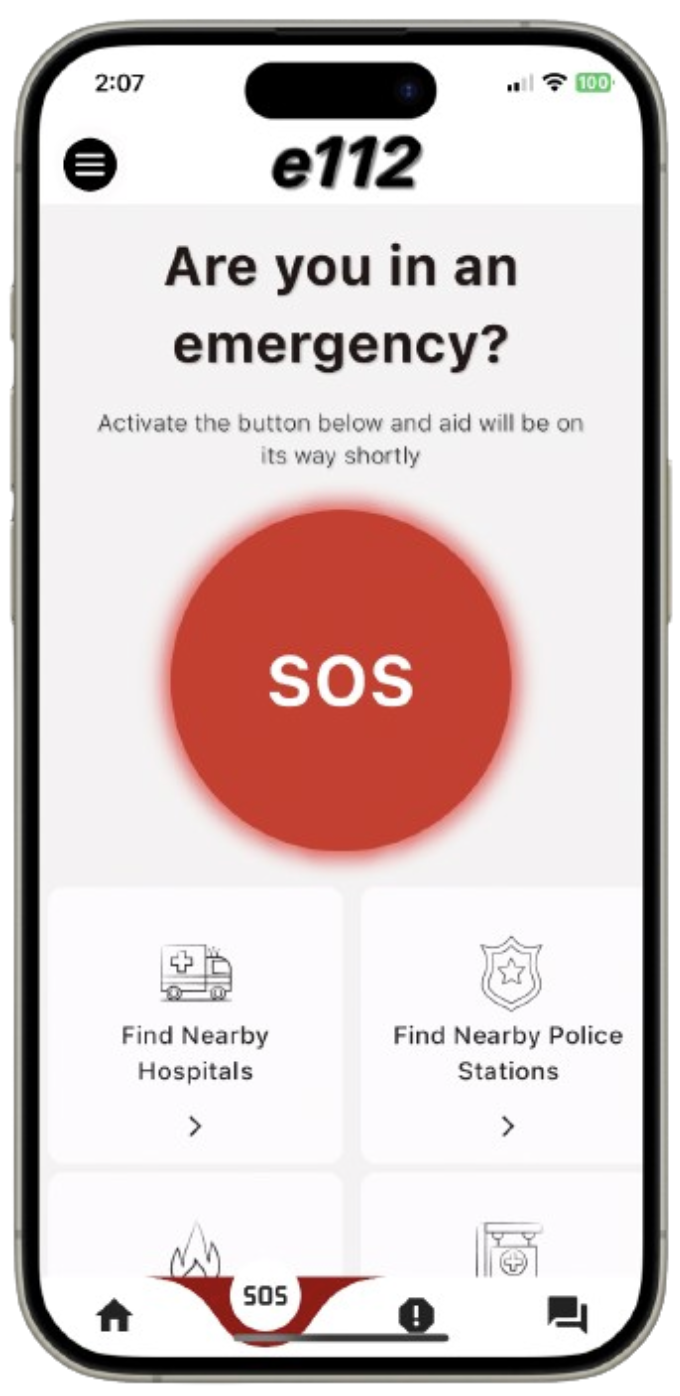}\\[-2pt]
    \footnotesize (e)
  \end{minipage}\hfill
  \begin{minipage}{0.32\linewidth}\centering
    \includegraphics[width=\linewidth]{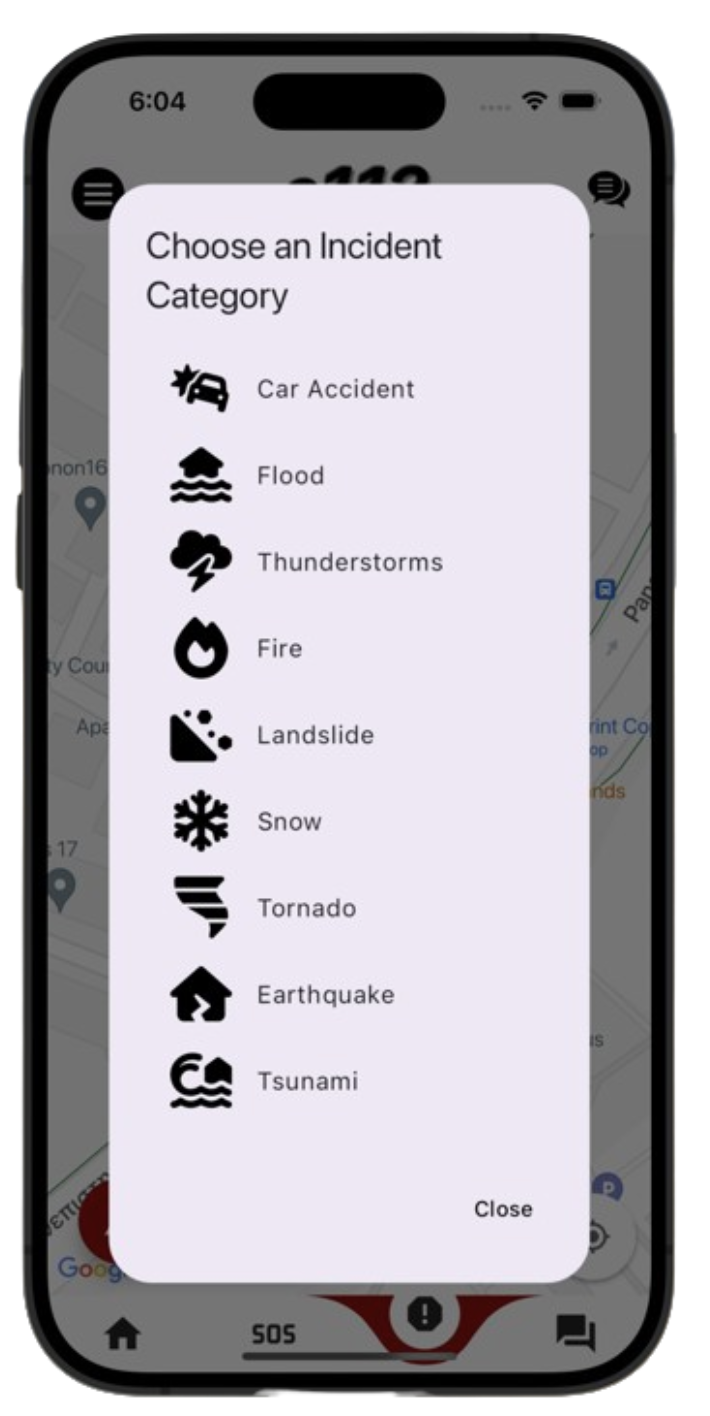}\\[-2pt]
    \footnotesize (f)
  \end{minipage}
  \caption{e112 mobile app functionality and interfaces.}
  \label{fig:functionality:1}
\end{figure}

\begin{figure}[t]
\centering
  \begin{minipage}{0.32\linewidth}\centering
    \includegraphics[width=\linewidth]{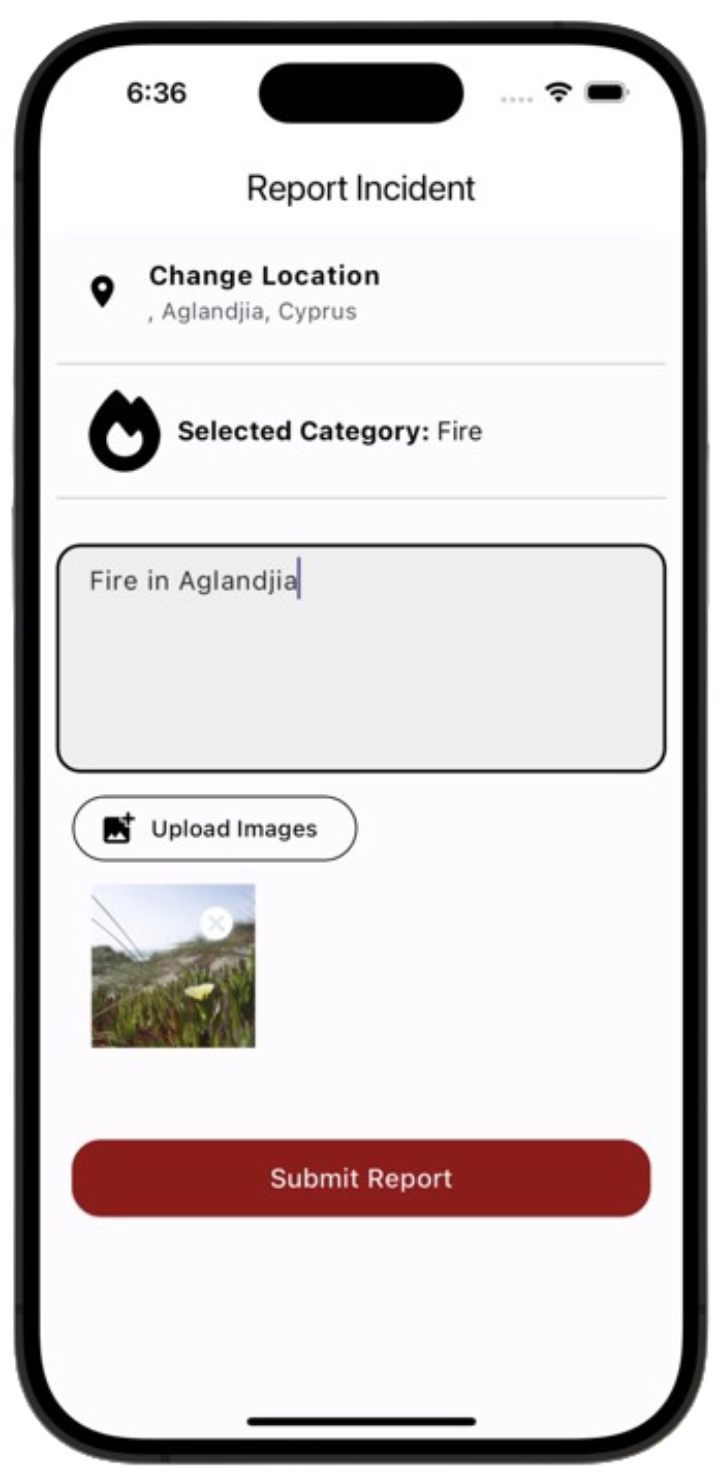}\\[-2pt]
    \footnotesize (g)
  \end{minipage}\hfill
  \begin{minipage}{0.32\linewidth}\centering
    \includegraphics[width=\linewidth]{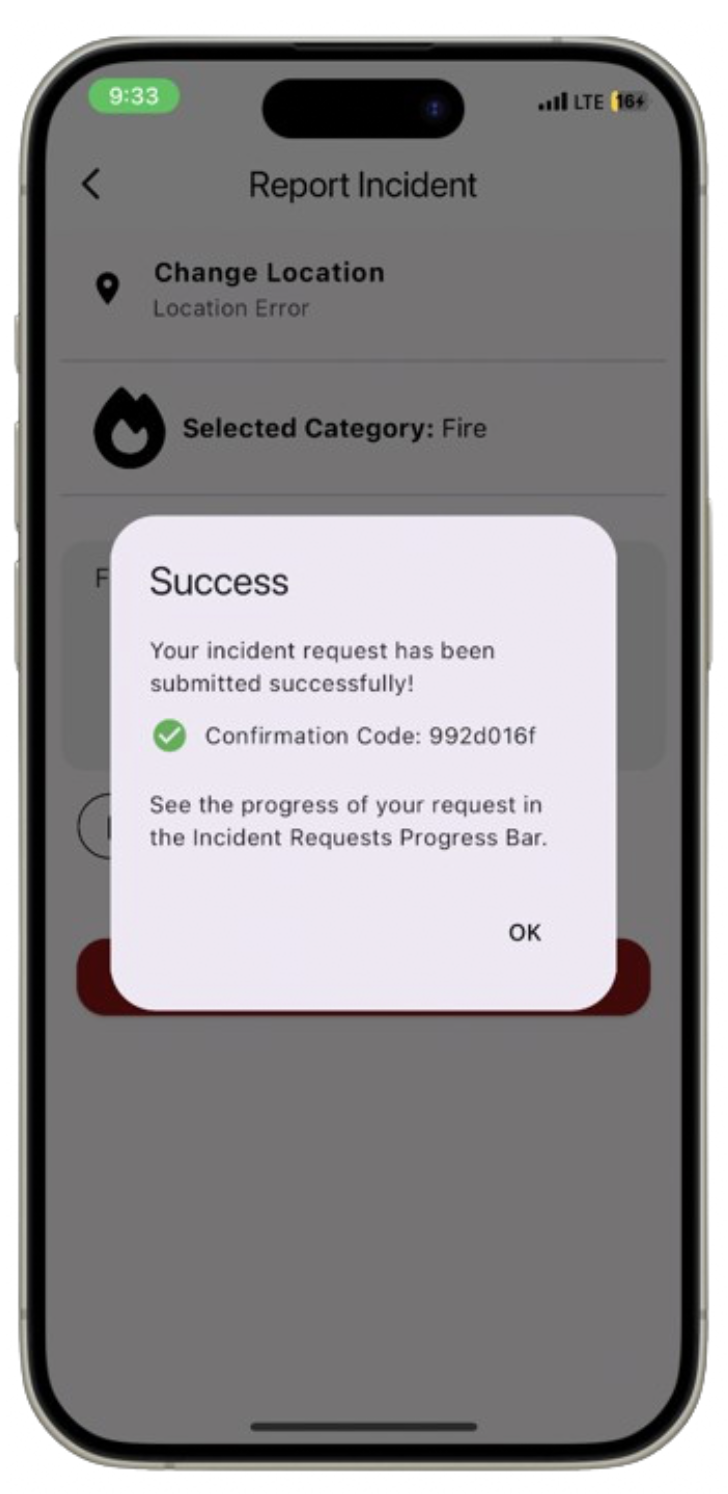}\\[-2pt]
    \footnotesize (h)
  \end{minipage}\hfill
  \begin{minipage}{0.32\linewidth}\centering
    \includegraphics[width=\linewidth]{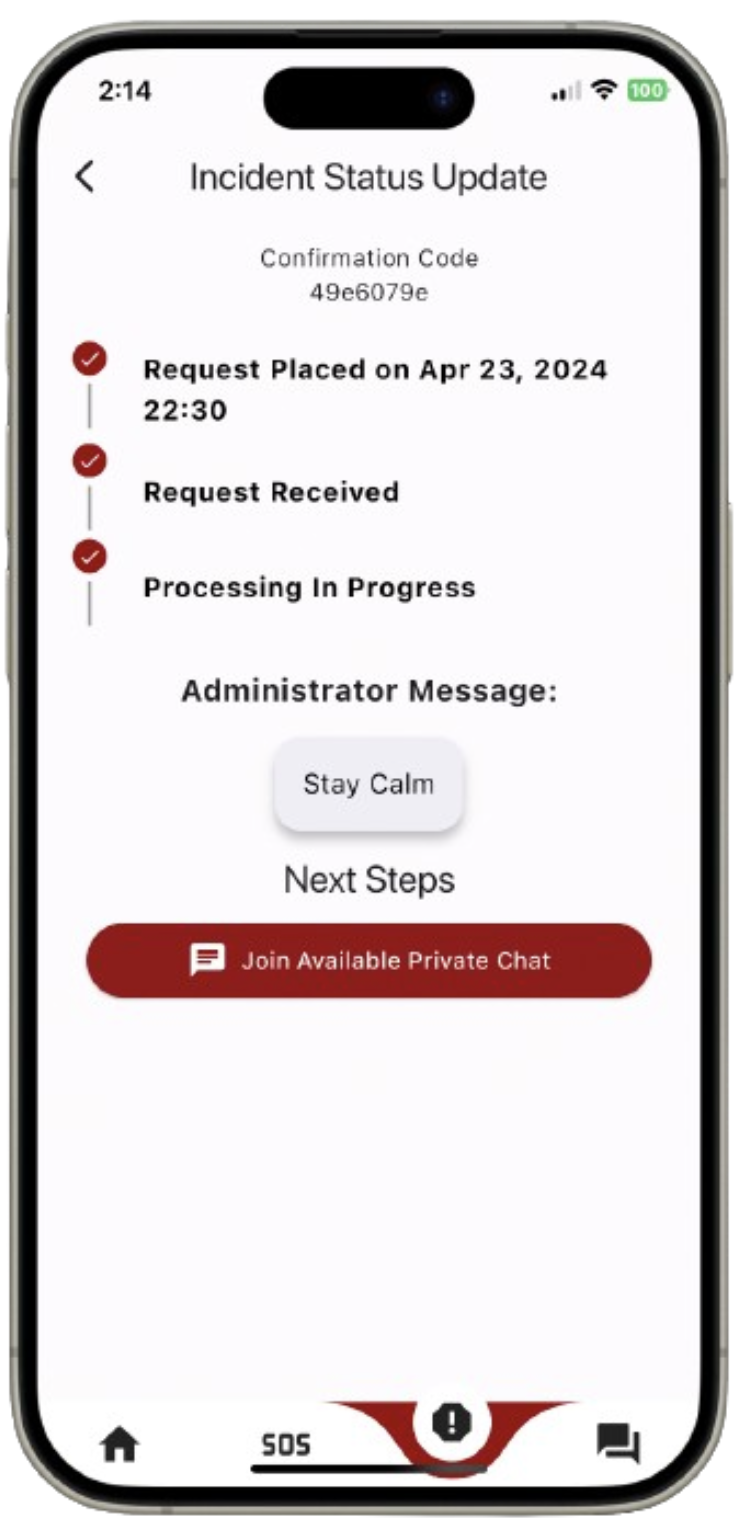}\\[-2pt]
    \footnotesize (i)
  \end{minipage}
  \begin{minipage}{0.32\linewidth}\centering
    \includegraphics[width=\linewidth]{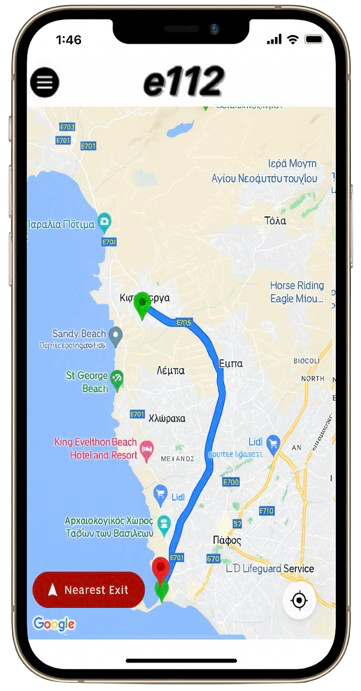}\\[-2pt]
    \footnotesize (j)
  \end{minipage}\hfill
  \begin{minipage}{0.32\linewidth}\centering
    \includegraphics[width=\linewidth]{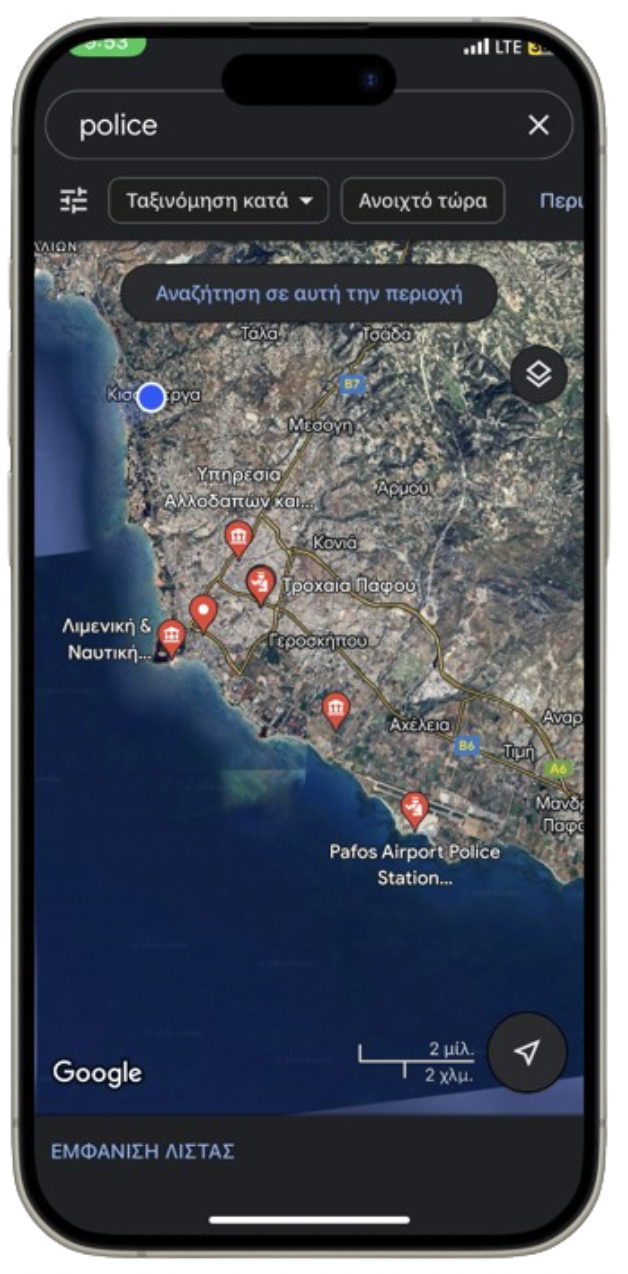}\\[-2pt]
    \footnotesize (k)
  \end{minipage}\hfill
  \begin{minipage}{0.32\linewidth}\centering
    \includegraphics[width=\linewidth]{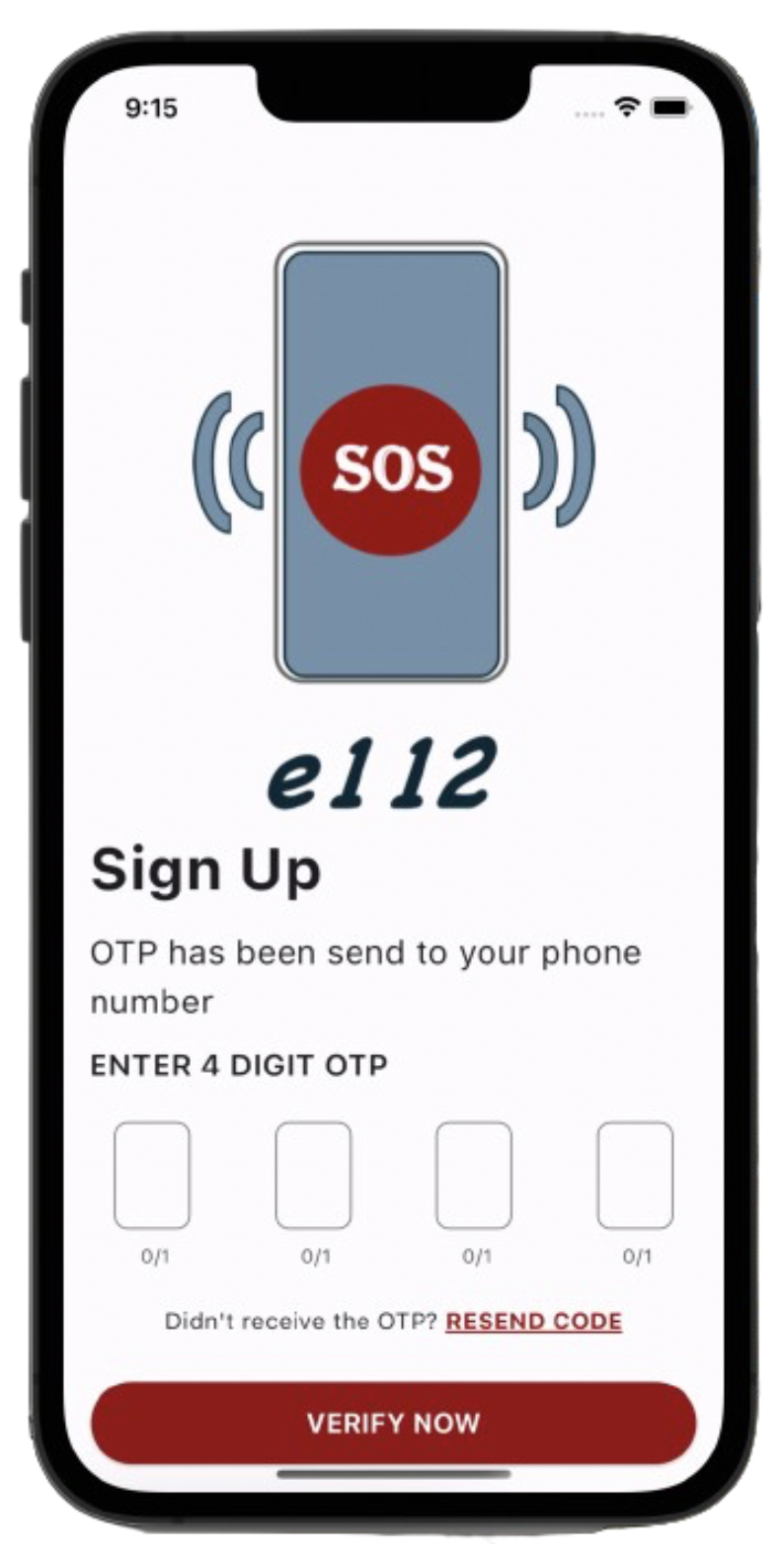}\\[-2pt]
    \footnotesize (l)
  \end{minipage}
  \caption{e112 mobile app functionality and interfaces (cont'd).}
  \label{fig:functionality:2}
\end{figure}

Through the UX analysis approach we established the following key functionalities
for the e112 mobile application:

\begin{itemize}[leftmargin=*, label=\textbullet]

\item \emph{Emergency Alerts' Dissemination}: The system delivers alerts and
situational updates in real time to people, based on their location, thereby
minimizing delays and reducing call-center overload during emergencies
(see Fig.~\ref{fig:functionality:1}a). Upon receipt of an alert, users are able
to immediately find simple instructions posted by emergency authorities on what
to do (see Fig.~\ref{fig:functionality:1}b,c) or seek help from people in the
area from a localized chat-group (see Fig.~\ref{fig:functionality:1}d).

\item \emph{Reporting to and Communication with Authorities}: The system enables
users to engage in trusted communication with authorities, requesting assistance
under personal duress (see Fig.~\ref{fig:functionality:1}e), reporting incidents
(see Fig.~\ref{fig:functionality:1}f, \ref{fig:functionality:2}g), getting
confirmation receipt (Fig.~\ref{fig:functionality:2}h), monitoring status
updates, and obtaining follow-ups (see Fig.~\ref{fig:functionality:2}i). The
design of the application's graphical user interface and interaction must adapt
to the user's anticipated level of stress, with a minimum number of clicks
required to ask for help when in personal danger (see
Figure~\ref{fig:functionality:1}e,f). To this end, information about the
location of the user and/or the reported emergency can be retrieved directly from
the smartphone and/or by using its GPS and camera (see
Figure~\ref{fig:functionality:2}g).

\item \emph{Information Resources and Presentation}: The application must present
information clearly and accurately in visual and textual formats, including
annotated maps and landmark images, while minimizing cognitive load under
high-stress conditions. The application shall provide intuitive navigation to
quickly locate critical resources such as safety guidelines, shelters, protected
spaces, hospitals, and police stations (see Fig.~\ref{fig:functionality:2}j,k).

\item \emph{Group Communication}: The system allows users affected by the same
emergency in the same location to join moderated chat groups that are dynamically
established by the platform's operator at the onset of an incident. Group chat
enables individuals to exchange information and advice, help each other even
before the arrival of rescue teams, and coordinate activities within a channel
monitored by emergency services (see Fig.~\ref{fig:functionality:1}d).
Moderation and monitoring will ensure that authorities maintain up-to-date
situational awareness, are able to deliver targeted guidance and coordinate
corrective action remotely, while protecting the integrity of the information
exchange from malicious information attacks.

\item \emph{Secure Accessibility}: The system facilitates registration through
verifiable, non-anonymous credentials (e.g., identifiable mobile numbers or
third-party authenticated accounts; see Fig.~\ref{fig:functionality:2}l) to
ensure accountability and prevent spam or disinformation attacks.

\item \emph{Interface Responsiveness}: The application adapts seamlessly across
devices and screen sizes to ensure a consistent user experience.

\end{itemize}

Additionally, for the graphical user interface (GUI) of the mobile application,
we adopted a color scheme chosen to balance urgency with usability under
stressful conditions, and facilitate navigation decisions under duress. The
primary color is dark red, symbolizing emergency and urgency, and used in
elements such as ``SOS'' and group chat buttons, to reinforce critical actions.
Red highlights affected areas on the evacuation map, while dark green is applied
to safe zones, providing an intuitive visual contrast that directs users toward
safety. Dark blue is reserved for basic guidelines, conveying reliability and
calmness, and dark yellow is used for evacuation points, combining visibility
with caution. By employing darker tones of each color, the design reduces visual
strain and enhances legibility, ensuring that users can quickly interpret
essential information even in high-stress disaster scenarios.

\section{System Architecture and Technology Stack}
\label{sec:arch}

\begin{figure*}[t]
\centering
\includegraphics[width=\textwidth]{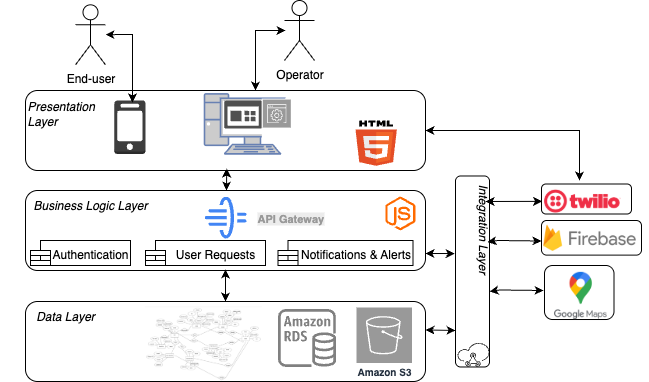}
\caption{e112 system architecture.}
\label{fig:arch}
\end{figure*}

The system architecture of e112 follows the principle of separation of concerns,
splitting the functionality into independent layers for presentation, business
logic, data management, and integration, so that each layer cannot affect another
and system modules can be implemented as cloud
micro-services~\citep{Trihinas2018c}. This makes the platform easier to scale
and troubleshoot.

\begin{itemize}[leftmargin=*, label=\textbullet]

\item The \emph{presentation layer} comprises the web dashboard used by e112
operators to log in, monitor incoming alerts, manage group communications, and
issue official updates (see Fig.~\ref{fig:dashb}), and the e112 application
which runs on end-user smartphones and provides reporting and communication
functionalities to citizens (see Fig.~\ref{fig:functionality:1}). Both dashboard
and app communicate with the back end through e112's API, which is implemented
in the business-logic layer. The e112 mobile application was implemented with
Flutter~\citep{Google:Flutter:2025}, a software framework for developing
native-quality mobile apps across Android and iOS from a single code-base.

\begin{figure}[h]
  \centering
  \begin{minipage}{0.90\linewidth}\centering
      \includegraphics[width=\linewidth]{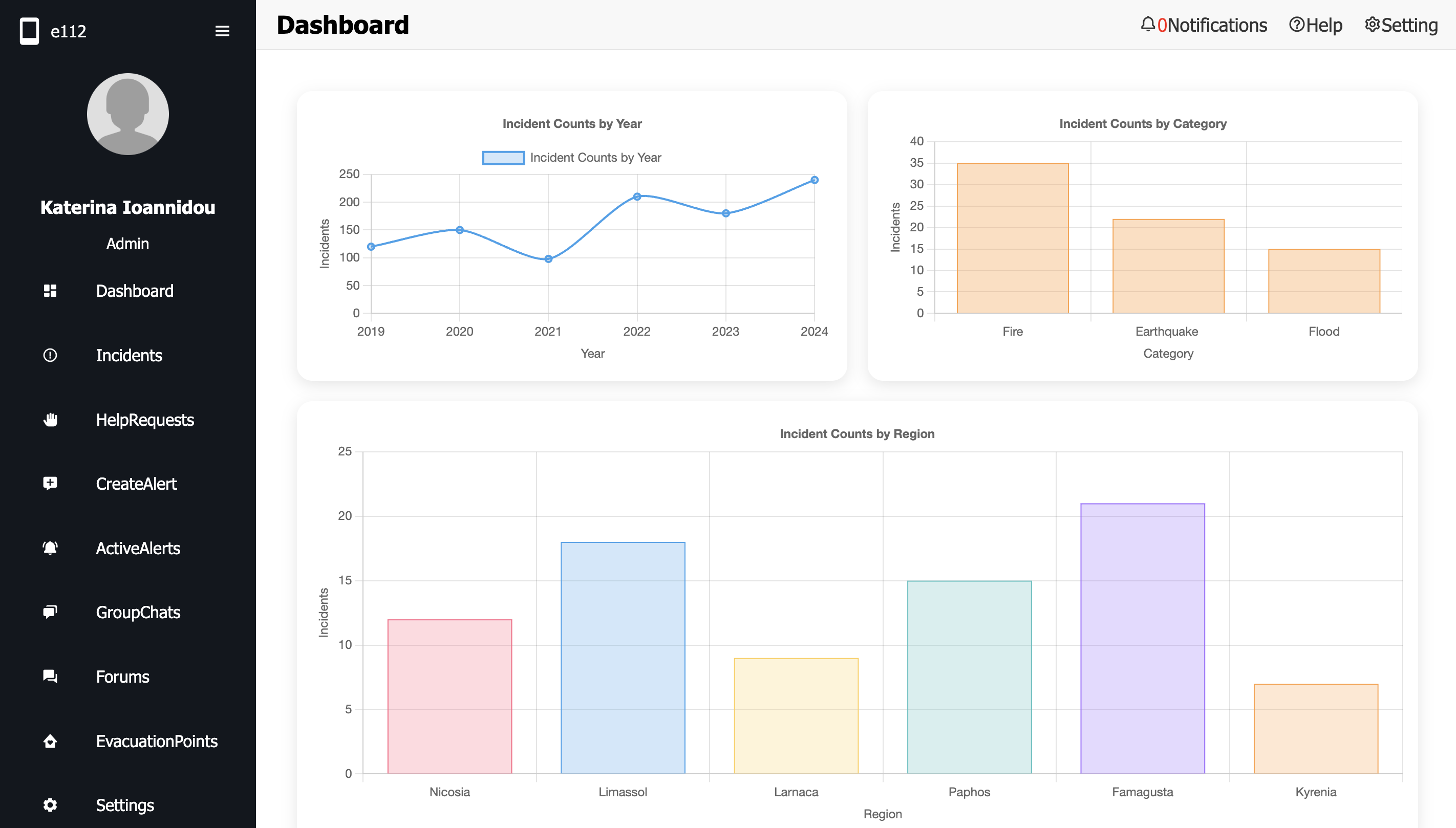}\\
    \includegraphics[width=\linewidth]{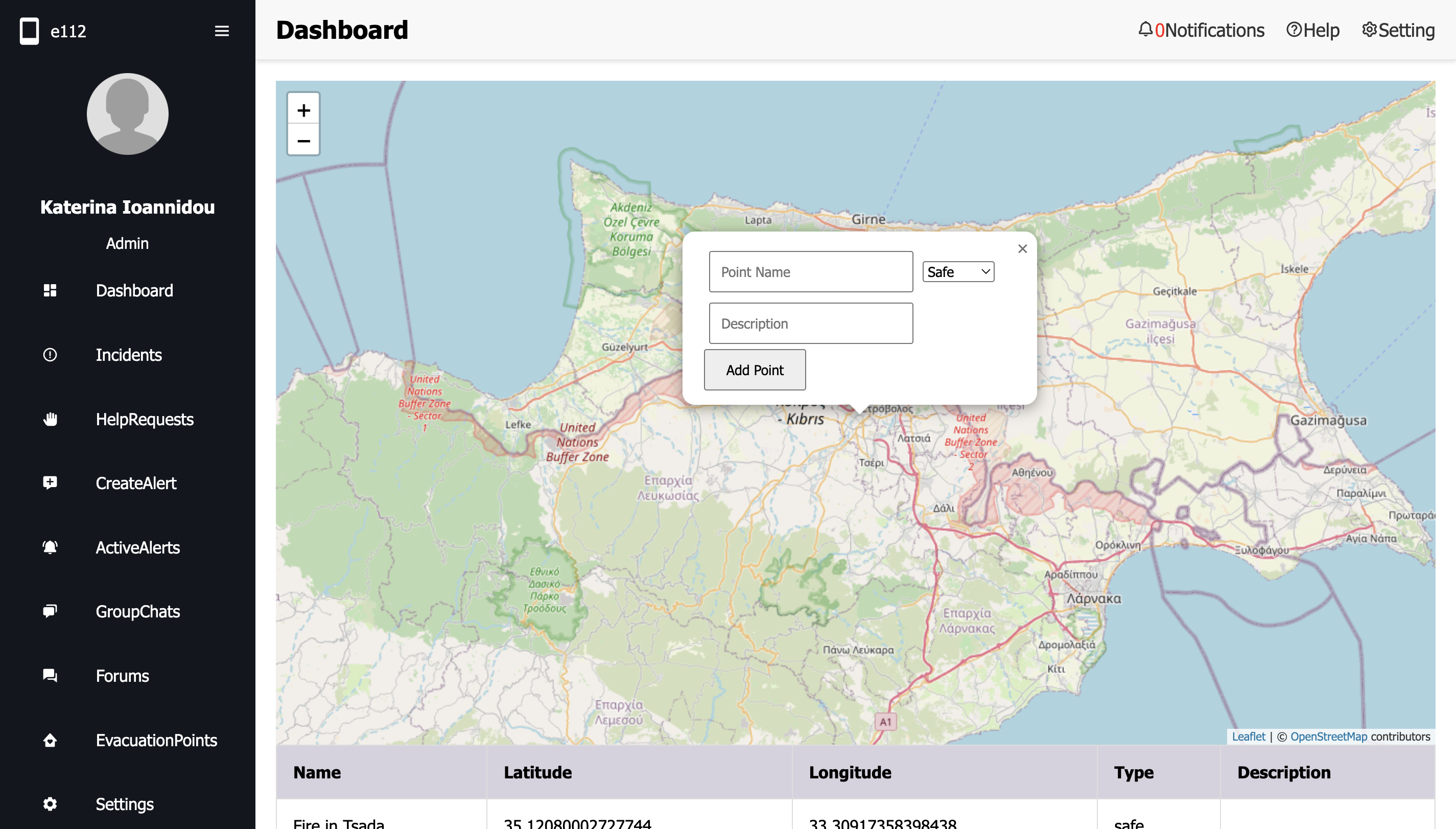}
  \end{minipage}
\caption{e112 administrator's dashboard: statistics and map-oriented interface.}
\label{fig:dashb}
\end{figure}

\item The \emph{business-logic layer} comprises the e112 API and is developed
with the Express-Node.js framework. This layer is responsible for managing the
main operations of the system including authentication, user requests, generation
and dissemination of alerts, and communication with the data layer. This layer
handles requests from both the mobile app and the operator's dashboard.

\item The \emph{data layer} is responsible for the persistent storage and
retrieval of all information required by the system to operate effectively. We
designed an entity--relationship model to capture core e112 entities and their
attributes, including users, alerts, incidents, help requests, chats, and
evacuation routes, along with their interrelationships. We implemented this model
as a relational PostgreSQL database deployed on Amazon RDS, ensuring reliability
and scalability for structured data. In parallel, we used Amazon S3 to store
multimedia files such as images and videos, providing durable and cost-effective
object storage. The data layer interacts with the business logic layer through
secure, private communication channels, ensuring both efficiency and data
integrity.

\item The \emph{integration layer} connects the business-logic layer with
external services, including the Google Maps API~\citep{GoogleMaps:2025} for
location information, the Twilio API~\citep{Twilio:Platform:2025} for mobile
phone verification through SMS, and the Firebase API~\citep{Firebase:2025} for
real-time push notifications. These services interact with the business logic
layer through RESTful HTTP endpoints.

\end{itemize}

We used the Vercel~\citep{Vercel:2025} Platform-as-a-Service (PaaS) cloud to
host the presentation and business-logic layers. Vercel provides serverless
deployment, automatic scaling, and global distribution.

\section{Evaluation}
\label{sec:eval}

The performance of an emergency application is determined not only by its
technical robustness but also by the clarity and efficiency of its user
experience, particularly under high-pressure
conditions~\citep{MentlerBerndtWesselHerczeg:2021}. To examine these dimensions
in e112, we evaluated its two main components: (i)~the administrator's dashboard,
through which authorized personnel receive, manage, and coordinate alerts, and
(ii)~the mobile application. To this end, we employed a mixed-methods evaluation
strategy, combining automated auditing of the administrative dashboard with
scenario-based usability testing of the mobile application. This approach was
selected to capture both quantitative performance indicators (e.g., page load
times, accessibility compliance) and qualitative insights into end-user
interaction under simulated emergency conditions.

\subsection{Administrator's Dashboard}

The administrative web interface was assessed using Unlighthouse, an open-source
auditing framework that reports on speed, accessibility, code quality, and
search engine optimization (SEO) metrics~\citep{Wilton:unlighthouse:2025}. Using
Unlighthouse, we measured the following Core Web Vitals:

\begin{itemize}[leftmargin=*, label=\textbullet]
    \item \emph{First Contentful Paint (FCP)}: the time taken to render the first
    visual element after page load.
    \item \emph{Largest Contentful Paint (LCP)}: the time at which the main
    content becomes visible.
    \item \emph{Total Blocking Time (TBT)}: the cumulative duration during which
    the main thread is blocked after FCP, delaying user interactions.
    \item \emph{Cumulative Layout Shift (CLS)}: the extent of unexpected visual
    shifts during page loading.
\end{itemize}

Each of these metrics is associated with a recommended threshold for an optimal
user experience. Specifically, an FCP $\le$1.8\,s is considered good, while
values $>$3\,s require improvement. LCP $\le$2.5\,s is desirable, with values
$>$4\,s negatively affecting usability. TBT should remain below 200\,ms to ensure
responsive interactions, while a CLS $<$0.1 is recommended to maintain visual
stability during loading. Based on our tests and measurements, the e112 dashboard
generally performed well. The recorded FCP slightly exceeded the recommended
threshold (1.8\,s) but remained within the acceptable range ($<$3\,s), possibly
reflecting the additional time required for font rendering. LCP was measured at
2.8\,s, marginally above the optimal threshold, likely due to the rendering of
large text blocks. In contrast, both TBT and CLS achieved ideal values,
demonstrating excellent responsiveness and visual stability.

The audit results further indicate a generally high-quality implementation of the
administrative dashboard. The system achieved a performance score of 90/100,
placing it within the ``good'' range of the Lighthouse
standard~\citep{Google:Lighthouse}. Accessibility achieved 95/100, consistent
with WCAG 2.1 guidelines~\citep{W3C:WCAG21}. The Best Practices score reached
96/100, confirming adherence to high standards of security and code
quality---both critical for systems that process sensitive emergency
data~\citep{ChromeBestPracticesDoctype}. By comparison, the SEO score was
82/100, an acceptable outcome but one that highlights the need for improvements
in metadata and indexing~\citep{LighthouseSEO}.

\subsection{Mobile Application}

For the e112 mobile application's usability study, we recruited 19 participants
from 16 to 65 years of age to account for known differences in technology use:
older adults often adapt more slowly to new interaction patterns whereas younger
users typically complete tasks more quickly and
efficiently~\citep{AgeEyeTracking2013,SONDEREGGER:2016}. This approach ensured
that the evaluation captured a representative spectrum of usability challenges
across diverse user demographics. All participants had prior experience with
smartphones and possessed basic digital literacy. For this study, we assumed a
simulated flooding disaster scenario to evaluate the application's usability in a
high-stress, real-world context. We asked participants to consider they are in
the midst of a flood disaster and use the e112 application to submit incident
reports and navigate the app's primary features. Following task completion,
participants filled a structured questionnaire measuring perceived usability,
task difficulty, and overall satisfaction.

We designed the questionnaire to collect meaningful and reliable feedback
capturing both quantitative ratings and qualitative perspectives. Our
questionnaire used a mix of five-point Likert scales, yes/no questions, and
short open-ended items. We asked participants to respond to 16 questions covering
a range of aspects of the application: ease of navigation and success in locating
desired information; the usefulness of alerts and real-time updates during an
emergency; which features they considered most valuable; whether they experienced
any difficulties during the reporting process; and how satisfied they were with
the response time. We also included items on the ease of posting in the forum,
the helpfulness of the evacuation-route tool, the utility of the guidelines
feature, and the value of the live chat function. Finally, we asked participants
how necessary they believed such an application would be in emergency situations.

After analyzing the responses, we found that participants generally considered
the application easy to navigate and reported no major difficulties with the user
interface. They also indicated high satisfaction with the response time. An
interesting result emerged regarding the live chat feature: participants aged
35--54 viewed it as unnecessary, while those aged 18--34 rated it as highly
useful. Overall, we observed high satisfaction with the application, with a mean
score of \emph{4.58 out of 5}, and a recommendation score averaging
\emph{4.63 out of 5}. Nearly all participants agreed that the existence of such
an application is important during emergency situations.

\section{Conclusions and Future Work}
\label{sec:concl}

The increasing frequency and scale of climate-driven disasters, often spanning
large geographic regions and populations, place significant pressure on Emergency
Services Communication Systems and demand a paradigm shift in how emergency
communications are designed and delivered. The widespread adoption of smartphones
offers a powerful opportunity to enable context-aware, two- and multi-way
interaction that leverages sensors, geo-location, and multimedia reporting. In
this paper, we presented the design, implementation, and evaluation of e112, a
cross-platform mobile application and cloud-based system supporting SOS requests,
incident reporting, evacuation guidance, targeted alerts, and moderated community
interaction, alongside an operator dashboard for situational awareness and
response coordination.

Our performance assessment and usability studies show that e112 is technically
robust, highly usable, and well received across age groups. A lightweight,
user-centered mobile platform can significantly strengthen preparedness and
response, reducing risk to human life during emergencies. Community-driven
features, such as contextualized chat forums, further enhance local coordination,
while the use of authenticated mobile identities and moderated channels reduces
the risk of misinformation attacks. Moderation can be further supported by
generative AI techniques. The platform's microservice-based architecture and
cloud-native implementation also enable scalable deployment, rapid iteration, and
seamless integration with third-party services.

Thanks to its modular design and use of HTTP/REST APIs, e112 is positioned to
interoperate with next-generation emergency communication standards such as
NG112~\citep{eena2024ltd} and NG911~\citep{wikipedia:NG911}. Both environments
rely on IP-native communication and have provisions for media transport, dynamic
location-to-service mapping, delivery of precise geo-location metadata, and
multi-modal ``total conversation'' capabilities (voice, text, image, video).

Several interesting open problems remain, including automated assessment of user
vulnerability, reliable integration with legacy emergency services, and the
management of high-volume, real-time communication during large-scale crises.
Future work focuses on integrating agentic systems to augment human operators by
classifying incoming requests, prioritizing incidents, monitoring group
communications, and filtering misinformation. Early results with large language
models (LLMs) show strong potential for automating report clustering,
urgent-case escalation, and conversation summarization, thereby improving
response times and enhancing overall emergency service effectiveness.

\section*{Use of AI Tools}

During the preparation of this work the authors used Nature's Research Assistant
(\url{https://researchassistant.nature.com}) and GPT-4 for searching for related
works and for improving the text. After using these tools, the authors reviewed
and edited the content as needed and take full responsibility for the content of
the published article.

\bibliographystyle{plainnat}
\bibliography{idrc2025-refs}

\end{document}